\documentclass[footinbib,twocolumn,showpacs,preprintnumbers,amsmath,amssymb,prl,superscriptaddress,longbibliography]{revtex4-1}

\usepackage{graphicx}
\usepackage{epsfig}
\usepackage{epstopdf}
\usepackage{dcolumn}
\usepackage{bm}
\usepackage{float}
\usepackage{color,xcolor}

\begin{document}
\title{
Modelling the Microstructure and Stress in Dense Suspensions \\ Under Inhomogeneous Flow} 
\author{J. J. J. Gillissen}
\email{jurriaangillissen@gmail.com}
\affiliation{
Department of Mathematics, 
University College London, 
Gower Street, 
London WC1E 6BT, 
United Kingdom}
\author{C. Ness}
\affiliation{
School of Engineering, 
University of Edinburgh,  
Edinburgh EH9 3FB, 
United Kingdom
}
\date{\today}
\begin{abstract}
Under inhomogeneous flow, dense suspensions exhibit complex behaviour that violates the conventional homogenous rheology. Specifically, one finds flowing regions with a macroscopic friction coefficient below the yielding criterion, and volume fraction above the jamming criterion. We demonstrate the underlying physics by incorporating shear rate fluctuations into a recently proposed tensor model for the microstructure and stress, and applying the model to an inhomogeneous flow problem. The model predictions agree qualitatively with particle-based simulations.
\end{abstract}
\maketitle

\paragraph{Introduction.--} 
Many materials, such as foods, cosmetics and ceramic precursors, consist of 
particles densely suspended in liquid, 
and their production relies on understanding the corresponding fluid mechanics~\cite{stickel2005fluid}. 
Despite a century of intense research including recent progress~\cite{baumgarten2019general}, 
comprehensive theoretical models are still lacking~\cite{denn2014rheology}.
Indeed, even the simple case of non-Brownian, non-inertial, hard spheres 
remains desceptively challenging~\cite{guazzelli2018rheology}. 
Under simple shear flow the mechanics are, in principle, governed by a single dimensionless parameter: 
specifying only the macroscopic friction coefficient $\mu=\Sigma_{xy}/\Pi$
sets the remaining non-dimensional variables, \emph{viz.}, the
volume fraction $\phi$ and the non-dimensional shear rate $S=\eta_s\dot{\gamma}/\Pi$~\cite{boyer2011unifying}.
Here $\Sigma_{xy}$ is the shear component of the 
stress tensor $\Sigma_{ij}$, 
$\Pi=-D^{-1}\Sigma_{ii}$ is the pressure (in $D$ dimensions),
$\eta_s$ is the viscosity of the suspending medium, 
and $\dot{\gamma}$ is the shear rate.
Carefully designed homogeneous flow experiments support this picture, revealing 
a decreasing $S$ and increasing $\phi$
upon reducing $\mu$, 
until the system jams ($S=0$) when $\phi$ reaches a maximum and 
$\mu$ reaches a minimum value. 
~\cite{boyer2011unifying}.
As opposed to frictionless particles that jam 
isotropically at random close packing $\phi=\phi_{\mathrm{RCP}}$ and $\mu=0$,
frictional particles jam with an anisotropic microstructure 
at $\phi=\phi_J$($<\phi_{\mathrm{RCP}}$) 
and $\mu=\mu_J>0$ \cite{wyart2014discontinuous}. 

Despite the conceptual power of this general result~\cite{guazzelli2018rheology}, its utility beyond homogeneous shear is limited.
In pressure driven Poiseuille flow, for example,
momentum conservation dictates that $\mu<\mu_J$ in a finite region around the centreline.
In this region, the [$S(\mu)$, $\phi(\mu)$] rheology described above clearly predicts jamming with $S=0$ and $\phi=\phi_J$.
This behaviour is not observed in experiments and particle-based simulations,
however,
which instead consistently show ``sub-yielding'' ($S>0$) and sometimes ``over-compaction'' ($\phi>\phi_J$),
in regions where $\mu<\mu_J$
\cite{
hampton1997migration,
lyon1998experimental, 
nott1994pressure, 
yeo2011numerical,
oh2015pressure}.
Making quantitative predictions of practical flows that comprise contiguous regions of $\mu>\mu_J$ and $\mu<\mu_J$ thus requires more detailed constitutive models that capture
both homogeneous rheology \emph{and}
the physics of
sub-yielding and over-compaction
that arise under inhomogeneous conditions. 
Although these effects have been addressed separately in the literature, there are no models available that capture both effects simultaneously.

Sub-yielding and over-compaction 
under inhomogeneous flow occur
in regions of vanishing shear rate,
where the
dynamics are completely governed by
fluctuating particle motions ~\cite{
miller2006normal,
isa2007shear,
kamrin2007stochastic, 
goyon2008spatial, 
bocquet2009kinetic, 
pouliquen2009non, 
kamrin2012nonlocal, 
bouzid2013nonlocal,
lecampion2014confined,
pahtz2019local}.
These fluctuations propagate from flowing regions 
with $\mu>\mu_J$ 
into (nearly solid) regions with $\mu<\mu_J$,
inducing particle rearrangements. 
This may allow the suspension to 
fluidise in otherwise solid regions,
with $\phi_J<\phi<\phi_\mathrm{RCP}$.

Attempts at incorporating over-compaction 
in constitutive models
are so far limited to 
linear extrapolation of the 
homogeneous $\phi(\mu)$ relation from regions with $\mu>\mu_J$ into regions with $\mu<\mu_J$~\cite{lecampion2014confined}.
The shape of the resulting density profiles, however, qualitatively differs from experimental data
\cite{hampton1997migration,lyon1998experimental,oh2015pressure}.
Sub-yielding, meanwhile, has been modelled 
by subjecting the fluidity (inverse viscosity) to a diffusion process \cite{kamrin2012nonlocal},
or by accounting for fluctuations in the expression for the suspension stress, with the fluctuation magnitude being computed 
using a transport equation borrowed from kinetic theory ~\cite{nott1994pressure}.
Alternatively, a simpler account for fluctuations can be 
derived by spatially 
averaging, i.e. filtering, the stress over 
a volume that is small compared 
to the system size and large compared to 
the particle size
\cite{mills1995rheology, morris1999curvilinear,miller2006normal}.
This 
leads to an increase in the normal viscosity but 
leaves the shear viscosity unaffected,
thereby reducing $\mu$ below $\mu_J$.
Crucially, these sub-yielding models
fail to account for 
microstructural changes due to fluctuations
and therefore do not capture over-compaction.

In this Letter we address these shortcomings, 
providing an intuitive explanation of sub-yielding and over-compaction. 
We do so by incorporating shear rate fluctuations into a recent microstructure model \cite{gillissen2018modeling,gillissen2019effect,gillissen2019taylor,gillissen2019constitutive,gillissen2020constitutive}.
When applied to inhomogeneous flows the resulting tensorial constitutive model 
predicts that fluctuations can: 
(i) isotropise the microstructure;
(ii) increase $\phi$ above $\phi_J$;
(iii) reduce $\mu$ below $\mu_J$. 
We compare the model predictions to those of particle-based simulations.

\paragraph{Constitutive model.--} The suspension stress tensor is modelled as \cite{gillissen2020constitutive}: 
\begin{equation}
\frac{\boldsymbol{\Sigma}}{\eta_s}=
2\langle\boldsymbol{E}\rangle+
\left[
\frac{\alpha_0\langle\boldsymbol{E}\rangle}{\left(1-\frac{\phi}{\phi_{\mathrm{RCP}}}\right)^2}+\frac{\chi_0 \langle\boldsymbol{E}_c\rangle}{\left(1-\frac{\xi}{\xi_J}\right)^2}
\right]
:\langle\boldsymbol{nnnn}\rangle.
\label{eq104}
\end{equation}
Here 
$\boldsymbol{n}$ is the separation unit vector of interacting particle pairs,
$\boldsymbol{L}=\boldsymbol{\nabla u}^T$ is the velocity gradient tensor and
$\boldsymbol{E}=\tfrac{1}{2}\left(\boldsymbol{L}+\boldsymbol{L}^T\right)$ is the rate of strain tensor,
which we decompose into
extensional $\boldsymbol{E}_e$ and compressive $\boldsymbol{E}_c$ parts:
\begin{equation}
\boldsymbol{E}_e=\tfrac{1}{2}\boldsymbol{E}+\tfrac{1}{4}
\vert\vert \boldsymbol{E}\vert\vert\boldsymbol{\delta},
\hspace{0.5cm}
\boldsymbol{E}_c=\tfrac{1}{2}\boldsymbol{E}-\tfrac{1}{4}
\vert\vert \boldsymbol{E}\vert\vert\boldsymbol{\delta}. 
\label{eq165}
\end{equation}
Note that Eq.~(\ref{eq165}) 
is valid 
in 2D but not in 3D.
Given the practical ubiquity of 2D shear we nonetheless proceed with Eq.~(\ref{eq165}).

In Eq. (\ref{eq104}) the filter operator $\langle\cdot\rangle$ averages over particle pairs that are contained in a space-time, \emph{filtering} volume 
which must be small compared to 
the spatial and temporal extents of the suspension and large compared to those of the fluctuations.
When carrying out the filtering, it has been assumed in Eq. (\ref{eq104}) [and in Eq. (\ref{eq106}) below] that 
the fluctuations in the velocity gradient field are uncorrelated with the fluctuations in the 
pair separation vector, e.g. 
$\langle\boldsymbol{E}_c\boldsymbol{nnnn}\rangle\approx 
\langle\boldsymbol{E}_c\rangle\langle\boldsymbol{nnnn}\rangle$.

In Eq. (\ref{eq104}) the jamming coordinate $\xi$ is defined as \cite{gillissen2020constitutive}:
\begin{equation}
\xi=-\frac{\langle\boldsymbol{nn}\rangle:\langle\boldsymbol{E}_c\rangle}{\sqrt{\langle\boldsymbol{E}_c\rangle:\langle\boldsymbol{E}_c}\rangle},
\end{equation}
which acts as a proxy for the coordination number $Z$, i.e. 
the number of direct contacts per particle~\cite{gillissen2019constitutive}.
The first and second terms in Eq. (\ref{eq104}) are, respectively, the stress induced by the fluid and by the particles. 
The latter contains lubrication and contact contributions, 
where 
$\alpha_0$ and $\chi_0$ are constants
and 
$\xi_J$ is the value of $\xi$ at
jamming.

In Eq. (\ref{eq104}) 
the fourth-order moment $\langle\boldsymbol{nnnn}\rangle$ of the orientation distribution function 
of $\boldsymbol{n}$ is expressed in terms of the second-order moment 
$\langle\boldsymbol{nn}\rangle$ using
\cite{hinch1976constitutive}:
\begin{multline}
\langle n_in_jn_kn_l\rangle=
-\langle n_mn_m\rangle\times\\
\frac{1}{(D+2)(D+4)}
\left(\delta_{ij}\delta_{kl}+\delta_{ik}\delta_{jl}+\delta_{il}\delta_{jk}\right)\\
+\frac{1}{D+4}\Big(
\delta_{ij}\langle {n}_k{n}_l\rangle+\delta_{ik}\langle {n}_j{n}_l\rangle+\delta_{il}\langle {n}_j{n}_k\rangle\\
+\langle {n}_i{n}_j\rangle\delta_{kl}+\langle {n}_i{n}_k\rangle\delta_{jl}+\langle {n}_i{n}_l\rangle\delta_{jk}
\Big).
\label{eq105}
\end{multline}
The second-order moment $\langle\boldsymbol{nn}\rangle$ is related to the
velocity gradient field with the following 
steady state balance equation 
\cite{gillissen2020constitutive}:
\begin{multline}
\boldsymbol{0}=
\langle{\boldsymbol{L}}\rangle\cdot\langle \boldsymbol{nn}\rangle+\langle \boldsymbol{nn}\rangle\cdot\langle{\boldsymbol{L}}^T\rangle-
2\langle{\boldsymbol{L}}\rangle:\langle \boldsymbol{nnnn}\rangle\\
-\beta\left[
\langle{\boldsymbol{E}}_e\rangle: \langle \boldsymbol{nnnn}\rangle+
\frac{
\color{black}
\phi
\color{black}
}{D(D+2)}
 \left(2\langle{\boldsymbol{E}}_c\rangle+\mathrm{Tr}(\langle{\boldsymbol{E}}_c\rangle)
 \boldsymbol{\delta}\right)
 \right].
  \label{eq106}
\end{multline}
The ``pair association rate'' $\beta$ controls the rate at which 
particle pairs are created and destroyed by fluid compression and extension, set respectively by 
$\boldsymbol{E}_c$ and 
$\boldsymbol{E}_e$. 
Eqs. (\ref{eq104}-\ref{eq106}) define a constitutive model for steady microstructure and stress in dense suspensions.

\paragraph{Incorporating fluctuations.--}
The shear rate consists of a mean $\dot{\gamma}=\vert\vert\langle\boldsymbol{E}\rangle\vert\vert=\sqrt{2\langle\boldsymbol{E}\rangle:\langle\boldsymbol{E}\rangle}$ and fluctuations.
While in homogeneous flow, 
the fluctuations are subdominant to the mean,
the fluctuations may dominate the mean in inhomogeneous flow, e.g. close to a Poiseuille centreline.
In those regions, although the filtered $\boldsymbol{E}$ is (nearly) zero, the 
filtered $\boldsymbol{E}_e$ and $\boldsymbol{E}_c$ are non-zero,
which is a consequence of the non-linearity of $\boldsymbol{E}_e$ and $\boldsymbol{E}_c$ in $\boldsymbol{E}$ [Eq. (\ref{eq165})]. 
Below we account for fluctuations 
in the model [Eqs. (\ref{eq104}-\ref{eq106})]
by filtering $\boldsymbol{E}_e$ and $\boldsymbol{E}_c$. 

In order to express $\langle \boldsymbol{E}_e\rangle$ 
and $\langle\boldsymbol{E}_c\rangle$
in terms of $\langle\boldsymbol{E}\rangle$,
we use that a fluctuating quantity $q=\langle{q}\rangle+q^{\prime}$ 
can be decomposed into its filtered $\langle{q}\rangle$ and its fluctuating  $q^{\prime}$
components,
and that $\langle q^{\prime}\rangle=0$. 
Filtering a non-linear function of $q$ gives additional terms. 
Specifically, filtering the absolute value of $q$ gives
$\langle\vert{\langle q\rangle}+q^{\prime}\vert\rangle 
\approx
\vert\langle q\rangle\vert+q_{\mathrm{rms}}$ \cite{note2}
where $q_{\mathrm{rms}}=\langle\vert q^{\prime}\vert\rangle$.
Similarly, filtering Eq. (\ref{eq165}) gives:
\begin{equation}
\langle\boldsymbol{E}_e\rangle =
\langle\boldsymbol{E}\rangle_e
+\tfrac{1}{4}\dot{\gamma}_{\mathrm{rms}}\boldsymbol{\delta},
\hspace{0.5cm}
\langle\boldsymbol{E}_c\rangle =
\langle\boldsymbol{E}\rangle_c
-\tfrac{1}{4}\dot{\gamma}_{\mathrm{rms}}\boldsymbol{\delta},
\label{eq162}
\end{equation}
where $\dot{\gamma}_{\mathrm{rms}}=\langle\vert\vert \boldsymbol{E}^{\prime}\vert\vert\rangle$ is 
the amplitude of the shear rate fluctuations.

In homogeneous shear flow, 
the fluctuating shear rate 
$\dot{\gamma}_{\mathrm{rms}}$ 
vanishes when the mean shear rate $\dot{\gamma}$ vanishes.
In inhomogeneous shear flows, on the other hand,
$\dot{\gamma}_{\mathrm{rms}}$ may remain finite when $\dot{\gamma}\rightarrow 0$, 
since fluctuations are propagating from nearby regions with finite $\dot{\gamma}$.
In this limiting case, the dynamics are dominated by 
$\dot{\gamma}_{\mathrm{rms}}$
and 
$\langle\boldsymbol{E}_e\rangle=-\langle\boldsymbol{E}_c\rangle=\tfrac{1}{4}\dot{\gamma}_{\mathrm{rms}}\boldsymbol{\delta}$.
Inserting these expressions into the filtered microstructure and stress equations [Eqs. (\ref{eq104}, \ref{eq105}, \ref{eq106})] 
gives isotropic tensors for the microstructure 
$\langle\boldsymbol{nn}\rangle\sim\boldsymbol{\delta}$
and the stress 
$\boldsymbol{\Sigma}\sim-\eta_s\dot{\gamma}_{\mathrm{rms}}\boldsymbol{\delta}$,
with negative normal stresses and zero shear stresses.
This behaviour corresponds to a vanishing macroscopic friction coefficient $\mu$, 
below the jamming friction coefficient for homogeneous systems $\mu_J$.
Our constitutive model similarly predicts isotropisation of the microstructure and stress in shear flow with superposed shear oscillations~\cite{gillissen2020constitutive}.

\begin{figure}
\includegraphics[width=1\linewidth]{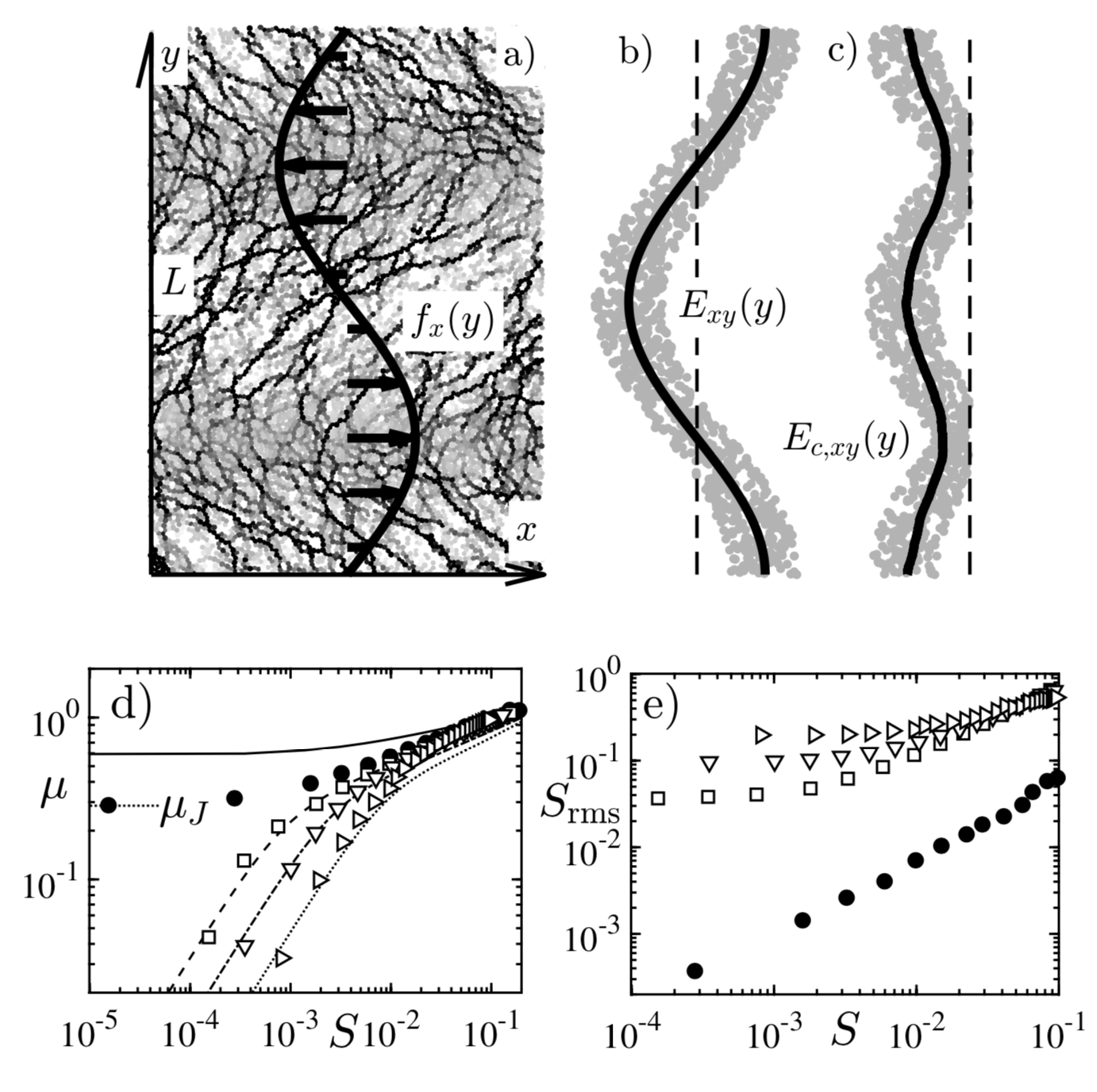}
\caption{
(a) Snapshot of particle-based simulation of Kolmogorov flow with 
relative domain size $L/a=278$.
The grey level  
indicates the particle pressure 
(black and white represent negative and positive, respectively)
and the driving force $f_x(y)$ is sketched with the black line.
(b, c) Sketches of the $xy$-component of ${\bm E}$ and ${\bm E}_c$ respectively, 
showing raw signals (grey) and filtered signals (black). 
The latter is zero at $y/L = 0.25$, $0.75$ for $\langle{\bm E}\rangle$ but not for $\langle{\bm E}_c\rangle$.
The dashed lines indicate 
the abscissas.
(d) Macroscopic friction coefficient $\mu$ as a function of the 
non-dimensional shear rate $S$  
predicted by simulation 
in homogeneous shear flow (filled circles) and
in Kolmogorov flow for 
$L/a=278$ (open squares), 139 (open downward triangles) and 56 (open rightward triangles)
and 
predicted by constitutive model for 
$S_{\mathrm{rms}}=0$ (solid line) $6\times 10^{-3}$ (dashed line), $1.5\times 10^{-2}$ (dash-dotted line) and 
$4.2\times 10^{-2}$ (dotted line).
(e) Non-dimensional shear rate fluctuations 
$S_{\mathrm{rms}}$ as a function of $S$  
predicted by simulation. The markers are as in (d). 
}
\label{fig1}
\end{figure}

\paragraph{Kolmogorov flow.--}
We apply the above model [Eqs. (\ref{eq104}-\ref{eq162})] to steady 2D Kolmogorov flow, driven by a body force density  
$\boldsymbol{f}=\hat{f}\sin(2\pi y/L)\boldsymbol{\delta}_x$ (with $\hat{f}$ the force amplitude)
pointing in the $x$-direction and oscillating in the $y$-direction with a period $L$ (Fig. \ref{fig1}a).
We chose this flow to test our model, as it is possibly the simplest inhomogeneous shear flow without solid surfaces.
In this inhomogeneous shear flow
$\boldsymbol{L}=\partial_y u_x\boldsymbol{\delta}_x\boldsymbol{\delta}_y$ 
and the fluid mechanical profiles are periodic in $y$ and independent of $x$ and $t$.
Figs. \ref{fig1}b-c show schematically the 
instantaneous and filtered  
profiles of the flow-gradient $xy$-component of the total deformation $\boldsymbol{E}$ and of its compressive part $\boldsymbol{E}_c$.
Crucially, the filtered $E_{xy}=0$ on the centrelines [at $y=L/4$ (mod $L/2$)], whereas the filtered $E_{c,xy} < 0$. 
This difference arises due to the non-linearity of $\boldsymbol{E}_c$ in $\boldsymbol{E}$ mentioned above, and demonstrates that fluctuations produce normal stresses 
but no shear stresses,  
resulting in sub-yielding close to the centrelines.

\paragraph{Particle-based simulation.--} We compare our constitutive model to particle-based simulations 
on 2D domains with dimensions in the $x$ and $y$-directions, respectively, of $L_x=200 a$ and 
$L=56 a$, $139 a$ and $278 a$. 
We use 
$N \sim 10^4$ 
bidisperse frictional spheres (radii $a$ and $1.4a$, stiffness $k$, density $\rho$)
and a domain averaged volume fraction:
\begin{equation}
\overline{\phi}=L^{-1}\int_0^{L}\phi(y)dy, 
\label{eq163}
\end{equation}
of $\overline{\phi}=0.7$.
The particles interact with each other through short-range
lubrication and frictional contact forces~\cite{cheal2018rheology} 
while drag forces between the particles and the suspending medium are 
omitted.
Instead, the flow is driven by a $y$-dependent force in the $x$-direction $f_0\sin(2\pi y/L)\boldsymbol{\delta}_x$ 
added to each particle. 
We set $f_0/ka=10^{-8}$, sufficiently small 
for the particles to behave as hard, inertia-free spheres  
($\rho\dot{\gamma} a^2/\eta_s<10^{-2}$).
The resulting driving force density is 
$\boldsymbol{f}=f_0n(y)\sin(2\pi y/L)\boldsymbol{\delta}_x$
where 
$n(y)$ is the particle number density and
the average force amplitude equals $\hat{f}=f_0\overline{n}=f_0N/(LL_x)$.
Simulations are run 
until a statistically steady state is reached in the entire domain 
and profiles are computed thereafter over $\dot{\gamma} t\approx 20$, 
based on the maximum $\dot{\gamma}$ in the domain.
We obtain velocity and structural profiles by averaging particle properties in $y$-bins, so that each single simulation provides a range of $\mu$ and $S$ values.

We also simulate 
2D homogeneous shear flow,  driven by Lees-Edwards boundary conditions, 
on a square domain with size $L=56 a$ and with $\phi=0.5-0.9$.
By measuring the divergence of the stresses with increasing $\phi$, we found the jamming friction coefficient to be $\mu_J=0.285$
and the limiting volume fractions as
$\phi_J=0.795$ and $\phi_{\mathrm{RCP}}=0.840$.

\paragraph{Model predictions of $\mu(S)$.--}
Fig. \ref{fig1}d shows the simulation results on ($S,\mu$)-coordinates 
under homogeneous shear and in Kolmogorov flow for various $L/a$. 
The data points correspond to fixed $\phi$ values in homogeneous shear 
and to fixed $y$-coordinates in the Kolmogorov simulation.
The inhomogeneous Kolmogorov flow simulation predicts sub-yielding, i.e.
$S>0$ in regions where $\mu<\mu_J$
while the homogeneous shear simulation predicts the homogenous $\mu(S)$ rheology consistent with Ref. \cite{boyer2011unifying}.

Shown in Fig. \ref{fig1}e are the simulated, non-dimensional shear rate fluctuations
$S_{\mathrm{rms}}={\eta_s\dot{\gamma}_{\mathrm{rms}}}/{\Pi}$
as a function of $S$ for the same cases as in Fig. \ref{fig1}d.
The shear rate fluctuations $\dot{\gamma}_{\mathrm{rms}}=\langle\vert\partial_yu_x^{\prime}\vert\rangle$ 
are calculated based on instantaneous, local realisations of 
$\partial_y u_x$, computed by fitting a linear function to the 
spatial distribution of the instantaneous particle velocities in a box of size $6a$.
The data show an increase 
in $S_{\mathrm{rms}}$ with a decrease in $L/a$
(that is, for steeper gradients of the driving force)
and a (non)vanishing 
$S_{\mathrm{rms}}$ in the limit of $S\rightarrow 0$ 
for the (in)homogeneous shear flow.
 
Constitutive model predictions are plotted with lines in Fig. \ref{fig1}d,
with $\alpha_0=\chi_0=0.96$, 
$\xi_J=0.6$
and $\beta=4$.
The latter two are not fitting parameters \emph{per se}, but follow from $\phi_{\mathrm{RCP}}=0.840$ and $\phi_J=0.795$~\cite{Note}.
Each line is obtained by solving $\phi$ and $S$ 
from Eqs. (\ref{eq104}-\ref{eq162}) 
for various values of $\mu$ at fixed $S_{\mathrm{rms}}$.
$S_{\mathrm{rms}}$ values
are chosen to best match the simulation data in Fig.~\ref{fig1}d (markers).
They are somewhat smaller 
than $S_{\mathrm{rms}}$ predicted by simulation (Fig. \ref{fig1}e),
reflecting that 
the constitutive model does not capture the correct quantitative relationship between 
$\mu$, $S$ and $S_{\mathrm{rms}}$.
Nevertheless,
the model predicts the correct qualitative behaviour,
specifically
$S_{\mathrm{rms}}>0$ 
results in sub-yielding, i.e. $\mu\to0$
as $S\rightarrow 0$,
with the effect being amplified as $S_\mathrm{rms}$ is increased.

\begin{figure}
\includegraphics[width=1\linewidth]{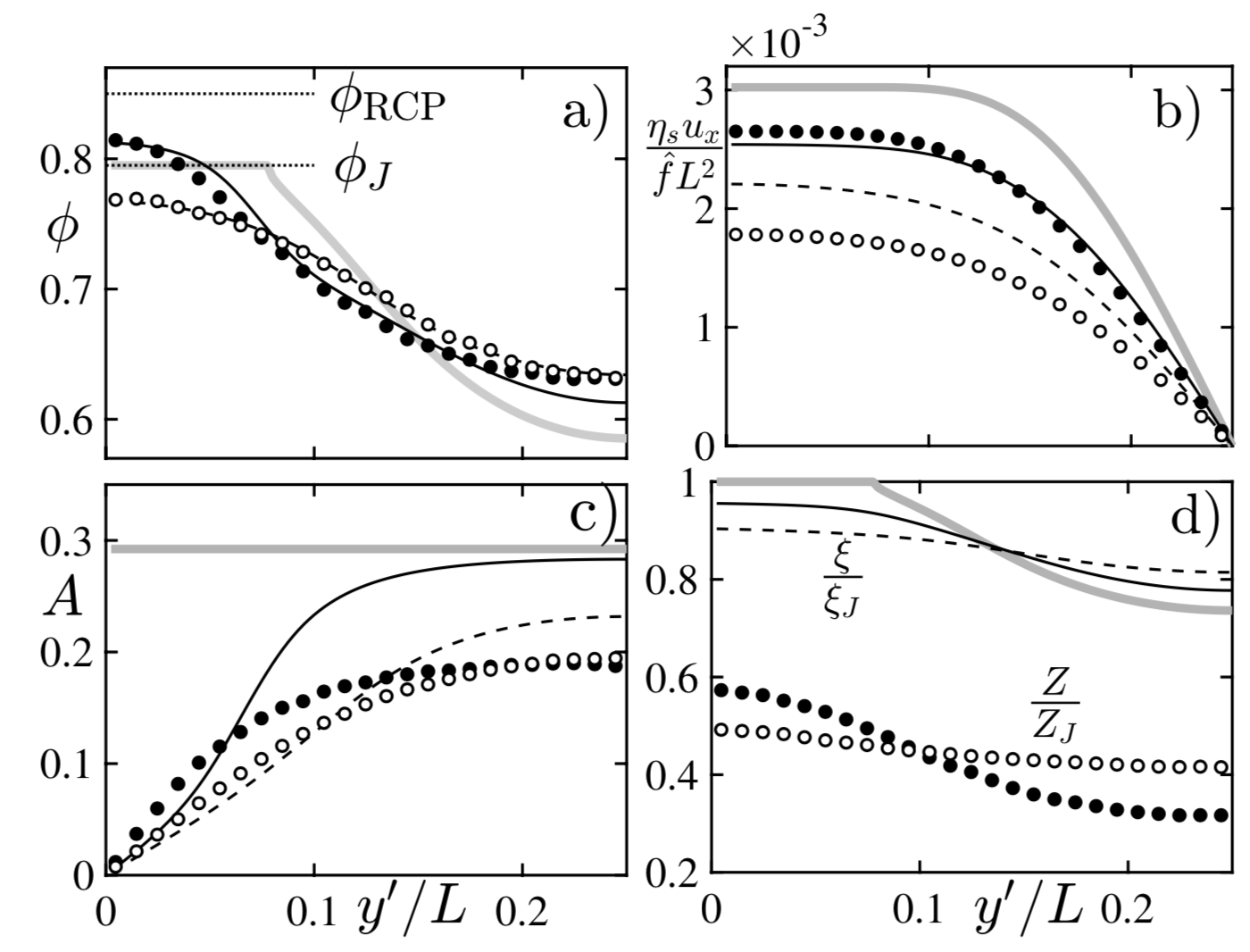}
\caption{
Kolmogorov flow profiles
as a function of the 
normalised 
distance to the nearest centreline $y^{\prime}/L$,
predicted by 
constitutive model with
$S_{\mathrm{rms}}=0$ (grey lines)
$S_{\mathrm{rms}}=10^{-2}$ (solid black lines) and $S_{\mathrm{rms}}=10^{-1}$ (dashed black lines)
and by
simulations with
$L/a=278$ (filled circles) and $56$ (open circles). 
(a) Volume fraction $\phi$;
(b) Normalised velocity $u_x/(\hat{f}L^2/\eta_s)$;
(c) Anisotropy of particle contacts $A$;
(d) Normalised coordination number $Z/Z_J$ in simulation and 
normalised jamming coordinate $\xi/\xi_J$ in constitutive model.
}
\label{fig2}
\end{figure}

\paragraph{Model predictions of profiles.--} Next we make predictions of the velocity and structural profiles in Kolmogorov flow by 
combining our constitutive model 
[Eqs. (\ref{eq104}-\ref{eq162})]
with the (inertia-free)
momentum balance 
$\boldsymbol{\nabla}\cdot\boldsymbol{\Sigma}+\boldsymbol{f}=\boldsymbol{0}$,
whose $x$ and $y$-components reduce to:
\begin{equation}
\Sigma_{xy}=\frac{\hat{f}L}{2\pi}\cos(2\pi y/L),\hspace{1cm}\Sigma_{yy}=-\mathrm{constant}.
\label{eq101}
\end{equation}
We use 
three non-dimensional shear rate fluctuations (assumed constant throughout the domain) 
$S_{\mathrm{rms}}=0$,
$10^{-2}$ and $10^{-1}$
where the 
former represents the homogeneous flow model and the
latter two are 
chosen to match the model to the simulated $\phi$-profiles in Fig. \ref{fig2}a (described below).
These $S_{\mathrm{rms}}$ values are different from those used in
Fig. \ref{fig1}d which were chosen to match the simulated $\mu(S)$ profiles. 
These differences again indicate the quantitative discrepancies between model and simulation.
We compute $\phi$, $\partial_yu_x$ and $\langle\boldsymbol{nn}\rangle$ 
in each $y$-coordinate for a given constant $\Sigma_{yy}$ 
from Eqs. (\ref{eq104}-\ref{eq162}, \ref{eq101}) 
using Newton-Raphson and then iteratively updating $\Sigma_{yy}$ 
using the bisection method
such that the 
integral volume fraction $\overline{\phi}$ [Eq. (\ref{eq163})] approaches $0.7$.

Shown in Fig. \ref{fig2} are profiles of 
the volume fraction $\phi$ (Fig. \ref{fig2}a), 
the non-dimensional suspension velocity $u_x\eta_s/(\hat{f}L^2)$ (Fig. \ref{fig2}b), 
the anisotropy $A$ of the particle contacts  
(Fig. \ref{fig2}c) 
and 
the coordination number $Z$
normalised by the value at homogeneous jamming $Z_J$
(Fig. \ref{fig2}d).
$Z$ is computed from the simulation output by counting contacting particle pairs (with $Z_J=3$), while $A$ is obtained by averaging $n_xn_y$ over all such pairs (with ${\bm n}$ the unit vector along the centre-to-centre line).
In the constitutive model, $Z/Z_J$ and $A$ 
are represented, respectively, 
by $\xi/\xi_J$ 
and 
$-({\langle\boldsymbol{E}_c\rangle:\langle\boldsymbol{nnnn}\rangle})_{xy}/{\langle\boldsymbol{nn}\rangle:\langle\boldsymbol{E}_c\rangle}$~\cite{gillissen2020constitutive}.
Due to symmetry Fig. \ref{fig2} only shows the profiles over one quarter of the wavelength $L$.

Without fluctuations, i.e. following the homogeneous rheology, the constitutive model predicts a jammed region around the centrelines
with $\dot{\gamma} = S=0$ and $\phi=\phi_J$ (grey lines in Fig. \ref{fig2}).
Fluctuations induce two effects. 
The first is an increase of the repulsive normal stress 
relative to the imposed shear stress, which is evidenced by a decrease in
$\mu$ for small $S$ in Fig. \ref{fig1}d.
This increased normal stress 
drives particles
away from the centrelines to the outer regions (Fig. \ref{fig2}a). 
In these outer regions  the shear rate is larger and the particles generate more shear stress than in the centreline regions.
This results in a lower non-dimensional velocity (Fig. \ref{fig2}b).
The second effect is isotropisation (i.e. $A\to0$) of the microstructure (Fig. \ref{fig2}c),
resulting in fewer particle contacts at a given $\phi$ (Fig. \ref{fig2}d).
This isotropisation allows $\phi$ to exceed $\phi_J$ 
and \textit{reduces} the normal stress near the centrelines.
These two competing effects may lead either to an increase in the volume fraction $\phi$ above 
$\phi_J$ (over-compaction, observed for $S_{\mathrm{rms}}=10^{-2}$) 
or to a reduction below $\phi_J$ (observed for $S_{\mathrm{rms}}=10^{-1}$) 
at the centrelines~(Fig. \ref{fig2}a).

Despite the qualitative agreement,
there are quantitative differences between the constitutive model and the particle-based simulation.
Fig. \ref{fig2}d for instance shows that $\xi/\xi_J$ in the constitutive model is larger 
than $Z/Z_J$ in the simulation.
There are many possible avenues for improving 
the quantitative accuracy of the model, e.g. by relaxing the assumption 
that 
velocity gradient fluctuations are uncorrelated with microstructure fluctuations
or by using complex relationships between 
the material functions $\alpha_0$, $\chi_0$ and $\beta$ 
and the state variables $\phi$, 
$\langle\boldsymbol{L}\rangle$ and $\langle\boldsymbol{nn}\rangle$. 
However, having demonstrated that our model contains a (possibly minimal) set of physics that can simultaneously reproduce sub-yielding and over-compaction,
we have chosen mathematical simplicity over quantitative accuracy, leaving the above developments as promising routes for further analysis. 

\paragraph{Conclusion.--} We have presented a tensorial model for the microstructure and stress in dense suspensions of frictional particles that includes the effect of fluctuations by applying a filtering to the 
microstructure balance equation.
In doing so, we are able to predict sub-yielding and over-compaction,
features common under practical flows but not predicted by homogeneous rheology models.


In addition to the potential model developments described above,
further improvements to the predictive capacity for practical applications 
will require testing in complex geometries.
We provide one such example in the Supplementary Material, 
namely a comparison between model and simulation predictions for pressure driven flow 
through a curved channel. 
Addressing the full details of this and other complex flows
will be the next step towards a
comprehensive fluid dynamical description of dense suspensions.

JJJG is supported by 
the Engineering and Physical Sciences Research Council of the United Kingdom 
Grant Number EP/N024915/1. 
CN acknowledges support from
the Royal Academy of Engineering under the Research
Fellowship scheme.
We thank H. J. Wilson, J. D. Peterson and M. E. Cates for stimulating discussions.

\appendix
\section{
Supplementary Information:\\
Comparison between constitutive model 
and discrete element method for flow through a curved channel}

\subsection{Flow Problem}
We consider a two dimensional (2D) flow of a dense suspension in a curved channel. 
The channel has an inner radius of $R_1$ and an outer radius of $R_2$.
The suspension flow is driven by a body force 
$\boldsymbol{f}=\hat{f} R_1r^{-1}\boldsymbol{\delta}_{\theta}$,
with $\hat{f}$ a constant. The body force points in the azimuthal $\theta$-direction. 
The volume averaged $\overline{\cdots}=2\int_{R_1}^{R_2}\cdots rdr/(R_2^2-R_1^2)$ body force equals 
$\overline{f}=2\hat{f}R_1(R_2-R_1)/(R_2^2-R_1^2)$.
No slip conditions are assumed on the walls. 
The flow is fully developed, 
i.e. the statistics of the flow only depend on the radial $r$-coordinate,
but not on time $t$ 
nor on the $\theta$-coordinate. 

\subsection{Constitutive Model}
The $\theta$-component of the momentum balance 
$\boldsymbol{\nabla}\cdot\boldsymbol{\Sigma}
=-\boldsymbol{f}$ reads:
$\partial_r\Sigma_{r\theta}+2r^{-1}\Sigma_{r\theta}=-\hat{f} R_1r^{-1}$,
which, after integration, gives:
\begin{equation}
\Sigma_{r\theta}=-\frac{\hat{f}}{2} + \frac{C_1 R_1}{r^2},
\label{eq001}
\end{equation}
where $C_1$ is an unknown, to be determined, integration constant.

The $r$-component of the momentum balance reads:
$\partial_r\Sigma_{rr}-r^{-1}(\Sigma_{\theta\theta}-\Sigma_{rr})=0$.
To simplify the analysis, we assume that 
$(\Sigma_{\theta\theta}-\Sigma_{rr})/\Sigma_{rr}$ is 
equal to an unknown, to be determined constant: 
\begin{equation}
C_2=\overline{\left(\frac{\Sigma_{\theta\theta}-\Sigma_{rr}}{\Sigma_{rr}}\right)}.
\label{eq005}
\end{equation} 
With this assumption, the $r$-momentum balance becomes:
$\partial_r\Sigma_{rr}-r^{-1}\Sigma_{rr}C_2=0$,
which, after integration, gives:
\begin{equation}
\Sigma_{rr}=-C_3\left(\frac{r}{R_1}\right)^{C_2},
\label{eq002}
\end{equation}
where $C_3$ is yet another unknown, to be determined, integration constant.

The overall particle volume fraction in the system is denoted $\overline{\phi}$, i.e.:
\begin{equation}
\overline{\phi}=\frac{2\int_{R_1}^{R_2}\phi(r)rdr}{R_2^2-R_1^2}.
\label{eq003}
\end{equation}

The velocity profile $U_{\phi}$ is related to the deformation rate $E_{r\phi}=(\partial_r-r^{-1})U_{\phi}$ through 
$U_{\phi}(r)=U_{\phi}(R_1)+r\int_{R_1}^{r}dr^{\prime}E_{r\phi}(r^{\prime})/r^{\prime}$.
Demanding that $U_{\phi}(R_1)=U_{\phi}(R_2)=0$ gives:
\begin{equation}
0=r\int_{R_1}^{R_2}\frac{E_{r\phi}(r^{\prime})}{r^{\prime}} dr^{\prime}.
\label{eq004}
\end{equation}

Provided $C_1$, $C_2$ and $C_3$,
we use Newton-Raphson (in an inner loop) 
to find $\phi$, $E_{r\phi}$ and $\langle\boldsymbol{nn}\rangle$ 
in each $r$ coordinate 
that satisfy Eqs. (\ref{eq001}, \ref{eq002}) and Eqs. (1-6) in the main text. 
We use an outer loop to update $C_2$ using Eq. (\ref{eq005}) and we use Newton Raphson 
to update $C_1$ and $C_2$  
in order to satisfy Eqs. (\ref{eq003}, \ref{eq004}).

We use the constitutive model to compute three flow cases with different radii of curvature. 
The corresponding radius ratios are 
$R_2/R_1=1, 1.5$ and $2$, respectively.
In order to have a good match between the velocity magnitude predicted by the constitutive model 
and the discrete element method (DEM)  
(described below) we use for these three cases
$\alpha_0=\chi_0=0.96, 1.15$ and 1.26, respectively. 
The other parameters in the constitutive model are
$\phi_{\mathrm{RCP}}=0.85$, 
$\phi_J=0.795$,
$\overline{\phi}=0.7$ and 
$S_{\mathrm{rms}}={E_{r\phi,\mathrm{rms}}}\eta_s/C_3=10^{-2}$.

\begin{figure}[H]
\includegraphics[width=1\linewidth]{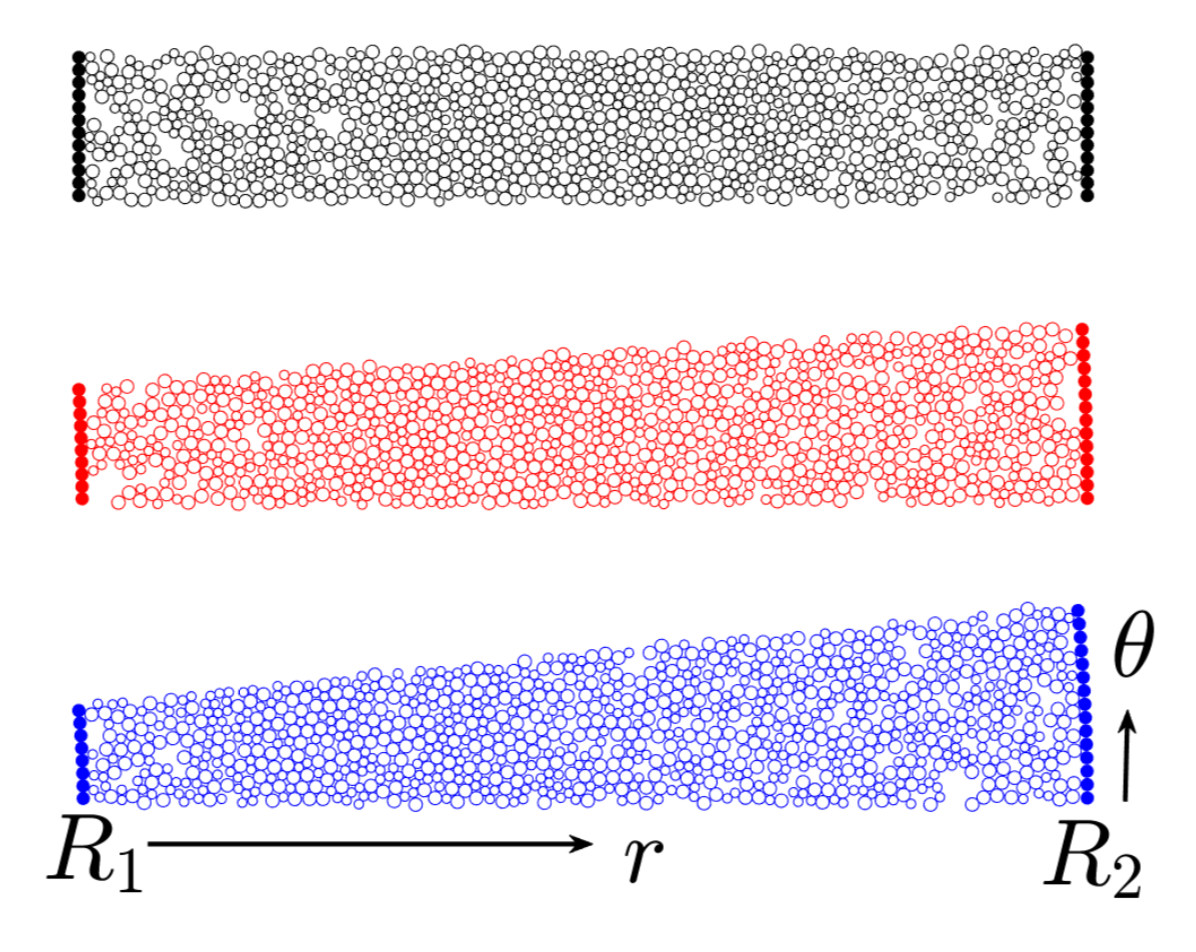}
\caption{
Particles in the DEM simulations with 
$R_2/R_1=1$ (top), 1.5 (middle) and 2 (bottom).
The empty particles are free and the filled particles are stationary and represent the solid boundaries.
}
\label{fig11}
\end{figure}

\subsection{Discrete Element method}
We compare the results from the constitutive model to those of a DEM.
For this purpose we use an in-house DEM code, 
that is similar to the one that is described in the main text.
In this code, solid walls are implemented as a collection of fixed particles and 
boundary conditions are implemented in the azimuthal direction that 
are periodic over an adjustable angle $\Delta \theta$. 

The flow in the DEM is driven by applying a force to each particle. 
This force points in the $\theta$-direction and it depends on the $r$-direction 
as $f_0(R_1/r)\boldsymbol{\delta}_{\theta}$. Here $f_0$ is a constant.
Assuming a homogeneous number density of particles $n=2N/[\Delta\theta(R_2^2-R_1^2)]$ this force distribution corresponds to a 
volume averaged body force density of $\overline{f}=2nf_0R_1(R_2-R_1)/(R_2^2-R_1^2)$.

We use 
$N=1000$ bi-disperse spheres with radii $a$ and 1.4$a$ and a radial  
domain size of $\Delta R/a=214$, with $\Delta R=R_2-R_1$.
We simulate three cases with $R_2/R_1=1, 1.5$ and 2 which correspond to a  
domain angle of $\Delta \theta= 0 ,0.06$ and $0.1$ 
and to a domain aspect ratio of $\Delta \theta R_1/\Delta R=6.7, 8.3$ and $10$, respectively.
The three computational domains with the simulated particles are illustrated in Fig. \ref{fig11}.

\subsection{Comparison}
\begin{figure}
\vspace{0.5cm}
\includegraphics[width=1\linewidth]{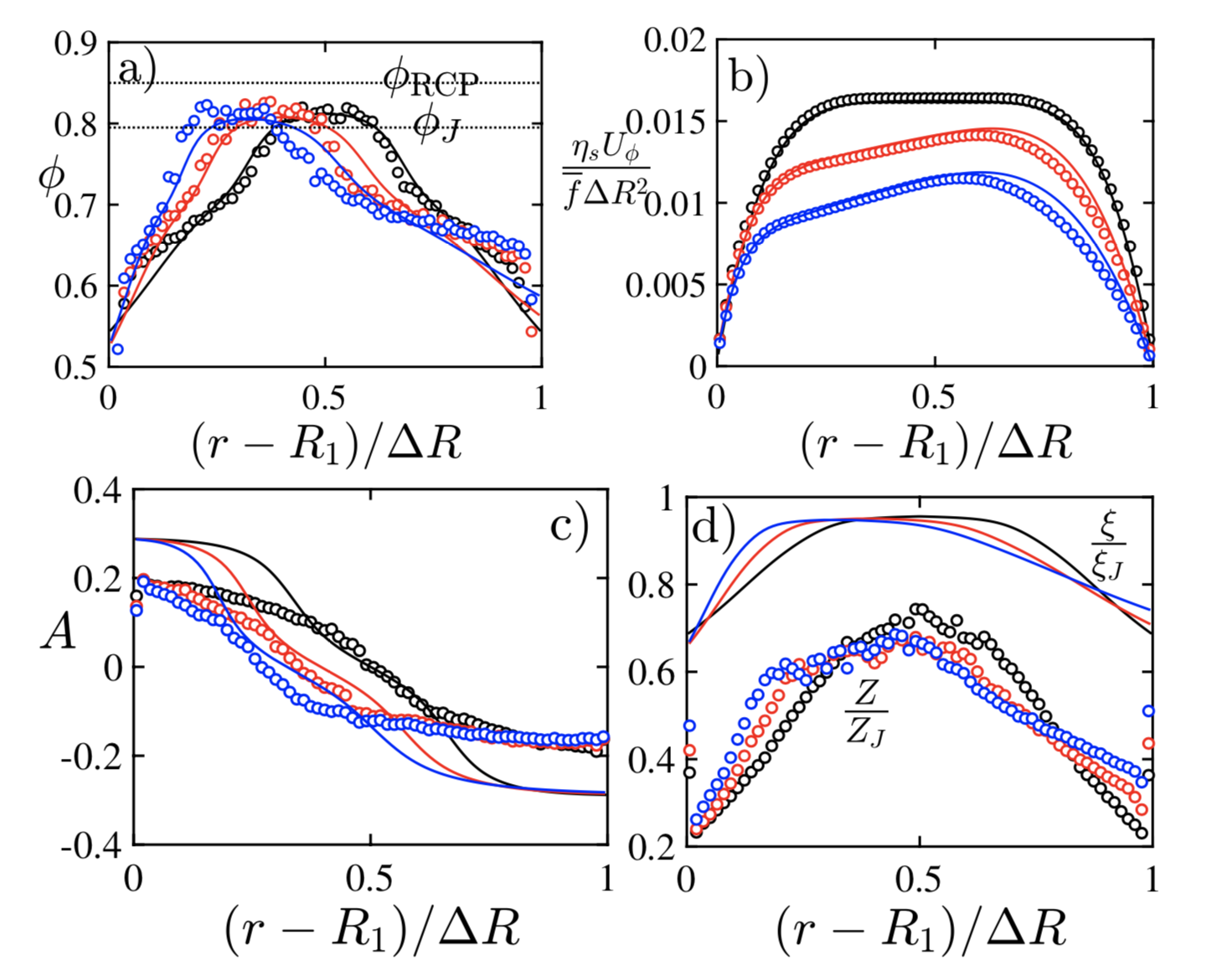}
\caption{
Comparison between the constitutive model (lines) and the DEM (markers) 
in terms of the profiles of the volume fraction (a), 
the suspension velocity (b), the contact microstructure anisotropy (c) and the coordination number (d) for 
$R_2/R_1=1$ (black), 1.5 (red) and 2 (blue).
}
\label{fig22}
\end{figure}

Fig. \ref{fig22} compares the constitutive model to the DEM in terms of the profiles of the volume fraction $\phi$, 
the non-dimensional suspension velocity $U_{\phi}\eta_s/(\overline{f}\Delta R^2)$, 
the contact microstructure anisotropy $A$ and the scaled coordination number, 
denoted $Z/Z_J$ in DEM and $\xi/\xi_J$ in the constitutive model. 
The definitions of these quantities are given in the main text.
Similar as in the Kolmogorov flow (described in the main text) 
the effects of flow inhomogeneity are concentrated in the region of vanishing shear stress. 
When the aspect ratio $R_2/R_1$ increases, 
this region moves from the channel center towards the inner wall.
This region is characterised by 
an isotropisation, i.e. $A\to 0$  (Fig. \ref{fig22}c)
and an over-compaction of the microstructure, i.e. $\phi>\phi_J$ (Fig. \ref{fig22}a).
As $R_2/R_1$ increases, 
the model also correctly captures an increase near the walls of the 
particle density (Fig. \ref{fig22}a) and the coordination number (Fig. \ref{fig22}d).
This redistribution results in a decrease in the suspension velocity (Fig. \ref{fig22}b).

\bibliography{article}

\begin{thebibliography}{32}
\bibitem [{not()}]{note2}%
  \BibitemOpen
  \href@noop {} {}\bibinfo {note} {When $\vert\langle q\rangle\vert\gg
  q_{\mathrm{rms}}$ we have $\langle\vert{\langle
  q\rangle}+q^{\prime}\vert\rangle \approx \vert\langle q\rangle\vert \approx
  \vert\langle q\rangle\vert+q_{\mathrm{rms}}$ and when $\vert\langle
  q\rangle\vert\ll q_{\mathrm{rms}}$ we have $\langle\vert{\langle
  q\rangle}+q^{\prime}\vert\rangle \approx q_{\mathrm{rms}} \approx
  \vert\langle q\rangle\vert+q_{\mathrm{rms}}$. For the intermediate case
  $\vert\langle q\rangle\vert\sim q_{\mathrm{rms}}$ we interpolate between
  these limiting cases, giving: $\langle\vert{\langle
  q\rangle}+q^{\prime}\vert\rangle \approx \vert\langle
  q\rangle\vert+q_{\mathrm{rms}}$.}\BibitemShut {Stop}%
\end{thebibliography}
\end{document}